\documentclass[epj]{webofc}
\usepackage[varg]{txfonts}  

\usepackage{hyperref}
\hypersetup{
  colorlinks=true,
  linkcolor=blue,
  citecolor=blue,
  urlcolor=blue
}

\wocname{EPJ Web of Conferences}

\woctitle{CONF12}

\begin{document}

\selectlanguage{english}
\title{Recent progress in anisotropic hydrodynamics}

\author{Michael Strickland}
\institute{Department of Physics, Kent State University, Kent, OH 44240 USA}

\abstract{
The quark-gluon plasma created in a relativistic heavy-ion collisions possesses a sizable pressure anisotropy in the local rest frame at very early times after the initial nuclear impact and this anisotropy only slowly relaxes as the system evolves.  In a kinetic theory picture, this translates into the existence of sizable momentum-space anisotropies in the underlying partonic distribution functions, $\langle p_L^2 \rangle \ll \langle p_T^2 \rangle$.  In such cases, it is better to reorganize the hydrodynamical expansion by taking into account momentum-space anisotropies at leading-order in the expansion instead of as a perturbative correction to an isotropic distribution.  The resulting anisotropic hydrodynamics framework has been shown to more accurately describe the dynamics of rapidly expanding systems such as the quark-gluon plasma.  In this proceedings contribution, I review the basic ideas of anisotropic hydrodynamics, recent progress, and present a few preliminary phenomenological predictions for identified particle spectra and elliptic flow.
}
\maketitle

\section{Introduction}
\label{intro}

The phenomenological application of viscous hydrodynamics to the dynamics of the quark-gluon plasma (QGP) created in heavy-ion collisions has been tremendously successful \cite{Romatschke:2009im,Gale:2013da,Jeon:2016uym}.  Despite this success, there are spacetime regions where standard viscous hydrodynamics is being pushed beyond its limits, such as at very early times after the initial heavy-ion collision $\tau < 1$ fm/c and near the (semi)-dilute edges of the system.  In these spacetime regions, viscous hydrodynamics itself tells you that there may be trouble, since the shear Knudsen number, $\text{Kn}_\pi = \tau_{\pi} \partial_\mu u^\mu$, and the inverse Reynolds number, $R_\pi^{-1} = \sqrt{\pi^{\mu\nu} \pi_{\mu\nu}}/P_{\rm eq}$, can become quite large.  The situation gets worse as the beam energy is decreased or one considers small collision systems such as pA and pp.  Additionally, when one wants to compute observables other than soft hadron production, such as heavy quarkonium suppression, photon emission, dilepton emission, etc. one traditionally employs a kinetic description which requires knowledge of the full momentum dependence of the underlying parton distribution function(s), $f(x,p)$.  In standard viscous hydrodynamics approaches, $f(x,p)$ is expressed as power series with the leading term being the isotropic thermal contribution and the shear and bulk corrections to isotropic equilibrium being expressed as polynomials in the momenta.  As a result, one cannot necessarily trust the high-momentum limit of production rates since this maps to regions in which the partonic distributions functions can become negative.

In order to address these problems,  in Refs.~\cite{Florkowski:2010cf,Martinez:2010sc} it was first shown that, in the context of relativistic transport theory, one could change the ansatz for the expansion point for the distribution function and use this reorganized expansion to derive so-called {\em anisotropic hydrodynamics} (aHydro).  In Ref.~\cite{Martinez:2010sc}, in particular, it was shown that, for a 0+1d boost invariant system, aHydro could reproduce both the ideal and free streaming limits of the dynamics and that, in the limit of small anisotropies, the dynamical equations reduced identically to those of Israel-Stewart viscous hydrodynamics.  Since these two original papers, there has been a great deal of progress in anisotropic hydrodynamics \cite{Florkowski:2010cf,Martinez:2010sc,Martinez:2012tu,Bazow:2013ifa,Tinti:2013vba,Nopoush:2014pfa,Tinti:2015xwa,Bazow:2015cha,Strickland:2015utc,Alqahtani:2015qja,Molnar:2016vvu,Molnar:2016gwq,Alqahtani:2016rth} including applications to cold atomic gases near the unitary limit \cite{Bluhm:2015raa,Bluhm:2015bzi}.  In parallel, there have been efforts to construct exact solutions to the Boltzmann equation in some simple cases which can be used to test the efficacy of various dissipative hydrodynamics approaches and it has been shown that anisotropic hydrodynamics most accurately reproduces all known exact solutions even in the limit of very large $\eta/s$ and/or initial momentum-space anisotropy \cite{Florkowski:2013lza,Florkowski:2013lya,Denicol:2014tha,Denicol:2014xca,Nopoush:2014qba,Heinz:2015gka,Molnar:2016gwq}.  The recent focus has been on turning aHydro into a practical phenomenological tool which has a realistic equation of state and self-consistent anisotropic hadronic freeze-out in order to make comparisons to experimental data.  In this proceedings, I review the moment formulation of aHydro which begins with an ansatz for the leading-order one-particle distribution that contains a symmetric anisotropy tensor which allows for multiple anisotropy parameters.  I present some preliminary comparisons with experimental data from LHC 2.76 TeV Pb-Pb collisions and provide an outlook for the future.

\section{Formalism}
\label{formalism}

\subsection{Distribution function ansatz}
\label{distfnc}

Typically, when one derives dissipative hydrodynamics from transport, the starting assumption is that the distribution function can be expanded around an assumed state of isotropic equilibrium with all non-equilibrium corrections collected into ``perturbative corrections'' which are typically expressed as an orthogonal polynomial expansion in the momentum contracted with the viscous stress tensor \cite{Denicol:2012cn,Denicol:2012es}.  In aHydro, one instead replaces the leading-order term in this series by an anisotropically deformed distribution function of the form~\cite{Martinez:2012tu,Tinti:2013vba,Nopoush:2014pfa}
\begin{equation}
f(x,p) \equiv f_{\rm eq}\left(\sqrt{p^\mu \Xi_{\mu\nu}(x) p^\nu}/\lambda(x) \right) \,,
\label{eq:dist}
\end{equation}
with $\lambda$ being a local temperature-like scale and $\Xi_{\mu\nu}\equiv u_\mu u_\nu+\xi_{\mu\nu}-\Phi \Delta_{\mu\nu}$ being the anisotropy tensor, which parametrizes the anisotropic form of distribution function.  In the definition of $\Xi^{\mu\nu}$, $u_\mu$ is fluid four-velocity, $\xi_{\mu\nu}$ is a traceless symmetric anisotropy tensor, $\Phi$ is related to the bulk-viscous pressure correction, and $\Delta_{\mu\nu}\equiv g_{\mu\nu}-u_\mu u_\nu$ projects out components of a four-vector which are transverse to $u^\mu$.  The quantities entering the ansatz obey $u^\mu u_\mu=1$, ${\xi^{\mu}}_\mu = 0$, and $u_\mu \xi^{\mu\nu} = u_\mu \Delta^{\mu\nu} = 0$.  Since $\xi^{\mu\nu}$ is transverse to $u^\mu$ it can be expanded in terms of spacelike basis vectors $X^\mu$, $Y^\mu$ and $Z^\mu$ which satisfy $u_\mu X^\mu = u_\mu Y^\mu = u_\mu Z^\mu = 0$.  The ansatz (\ref{eq:dist}) contains information about both shear and bulk corrections at leading order \cite{Nopoush:2014pfa}.  In what follows, we assume the distribution to be of Boltzmann form, i.e. $f_{\rm iso}(p)\equiv \exp{(-p)}$ and that the anisotropy tensor $\xi^{\mu\nu}$ is diagonal in the local rest frame (LRF)
\begin{equation}
\xi^{\mu\nu}(x) = \xi_x(x) X^\mu X^\nu + \xi_y(x) Y^\mu Y^\nu + \xi_z(x) Z^\mu Z^\nu \, .
\end{equation}
Since $\xi^{\mu\nu}$ is traceless, there are only two independent anisotropies.  In practice, it is convenient to parameterize these two independent degrees of freedom and the $\Phi$ parameter in terms of three momentum-space ellipticities
\begin{equation}
\alpha_i \equiv (1 + \xi_i + \Phi)^{-1/2} \, ,
\end{equation}
in which case, under the assumption that the anisotropy tensor is diagonal in the LRF, allows us to write simply
\begin{equation}
f(x,p)  = f_{\rm eq}\!\left(\frac{1}{\lambda(x)}\sqrt{\sum_i \frac{p_i^2}{\alpha_i^2(x)} + m^2}\right)  .
\label{eq:faniso}
\end{equation}

\subsection{Equations of motion}
\label{eqom}

In anisotropic hydrodynamics, the equations of motion are obtained from moments of the Boltzmann equation
\begin{equation}
p_\mu \partial^\mu f = - C[f] \, ,
\end{equation}
with the collisional kernel taken in the relaxation time approximation (RTA) 
\begin{equation}
C[f] = \frac{p^\mu u_\mu}{\tau_{\rm eq}(x)} [ f - f_{\rm eq}(|{\bf p}|,T(x)) ] \, .
\end{equation}
In what follows, we have used the first and second moments, which result in equations of the form
\begin{equation}
\begin{aligned}
&\partial_\mu T^{\mu\nu} = 0  \, , \\
&\partial_\mu I^{\mu\nu\lambda}  = \frac{u_\mu}{\tau_{\rm eq}} (I^{\mu\nu\lambda}_{\rm eq} - I^{\mu\nu\lambda}) \, ,
\end{aligned}
\end{equation}
where $T^{\mu\nu}$ is the energy-momentum tensor consistent with Eq.~(\ref{eq:faniso}) and $I^{\mu_1 \cdots \mu_n}$ is defined by
\begin{equation}
I^{\mu_1 \cdots \mu_n} \equiv N_{\rm dof} \int \frac{d^3 p}{(2\pi)^3} \frac{p^{\mu_1} \cdots p^{\mu_n} }{E} f(x,p)
\end{equation}
where $N_{\rm dof}$ is the number of degrees of freedom.  The general equilibrium tensor $I^{\mu_1 \cdots \mu_n}_{\rm eq}$ is the same with $ f(x,p) \rightarrow  f_{\rm eq}(|{\bf p}|,T(x))$.  In terms of the general moments, $I^{\mu_1 \cdots \mu_n}$, one has $N^\mu = I^\mu$, $T^{\mu\nu} = I^{\mu\nu}$, etc.  With the assumption that the anisotropy tensor is diagonal, we need equations of motion for the three ellipticities, three independent components of $u^\mu$, and $\lambda$.  We obtain these from the four equations for energy-momentum conservation and three diagonal projections of the second moment equation.  The local effective temperature $T(x)$ is determined by Landau-matching, which requires that the equilibrium and non-equilibrium energy densities are the same at all points in spacetime, i.e. $\varepsilon(\vec{\alpha},\lambda) = \varepsilon_{\rm eq}(T)$.

\subsection{Equation of State}
\label{eos}

If the QGP is anisotropic in momentum-space it is not obvious how one should implement a realistic lattice-based equation of state since this is an implicitly equilibrium concept.  There are currently two approaches being followed in the literature.  In the first, dubbed the {\em standard approach}, one obtains the dynamical equations necessary in the conformal limit, $m\rightarrow 0$. In this limit, the components of $T^{\mu\nu}$ multiplicatively factorize and one can then connect the isotropic parts of energy density and pressures using a realistic EoS \cite{Nopoush:2015yga}.  In the second approach, dubbed the {\em quasiparticle approach}, one assumes that the system is comprised of quasiparticles with a temperature-dependent mass, $m \rightarrow m(T)$, which is fit to lattice QCD data \cite{Alqahtani:2015qja,Alqahtani:2016rth}.  The quasiparticle method is more self-consistent in the way breaking of conformal invariance is implemented, however, at the moment it is infeasible to use for 3+1d simulations.  For this reason, in the results presented herein we will use the standard method for implementing the equation of state.  For the underlying realistic equilibrium EoS we used the Krakow parametrization of lattice data which is matched onto a hadron resonance gas at low temperatures~\cite{Chojnacki:2007jc}.

\subsection{Freezeout}
\label{freezeout}

As the system cools, the QGP undergoes a crossover from quark and gluons to hadronic degrees of freedom which subsequently freeze out kinetically when their mean free path becomes large.   In order to compare the result of hydrodynamic models to experimental data, one needs to calculate the differential particle spectra at freezeout. In practice, we construct a constant energy density hypersurface, defined through the effective temperature $T_{\rm FO}=\varepsilon^{-1}(\varepsilon_{\rm FO})$. Then, by computing the number of particles that cross this hypersurface, one can determine the number of hadrons produced in heavy-ion collisions at the freezeout using
\begin{equation}
\bigg(p^0\frac{dN}{d^3p}\bigg)_{i}=\frac{{\cal N}_i}{(2\pi)^3}\int \! f_i(x,p) \, p^\mu d^3\Sigma_\mu \, ,
\label{eq:dNdp3}
\end{equation}
where $i$ labels the hadronic species, ${\cal N}_i\equiv(2s_i+1)(2g_i+1)$ is the degeneracy factor with $s_i$ and $g_i$ being the spin and isospin of the state in question, and $f_i$ is the distribution function for particle species $i$ taking into account the appropriate quantum statistics.  For each species, we assume that the same anisotropic distribution function form can be used \cite{Nopoush:2015yga,Alqahtani:2016rth}.

\section{Computational setup}
\label{compsetup}

\subsection{Initial Conditions}
\label{init}

In order to solve the aHydro dynamical equations one has to choose initial conditions at $\tau = \tau_0$, i.e. for 3+1d aHydro with only diagonal anisotropies in the LRF, one needs seven three-dimensional profiles: $\lambda(\tau_0, {\bf x_\perp}, \varsigma)$, $\vec{\alpha}(\tau_0, {\bf x_\perp}, \varsigma)$, and $\vec{u} (\tau_0, {\bf x_\perp}, \varsigma)$, where $\varsigma$ is the spatial rapidity.  In this work, we assume that the initial transverse profile for the effective temperature (determined via Landau matching) is given by the optical Glauber model.  We assume that the initial energy density is proportional to the scaled initial density of the sources such that the initial effective temperature is
\begin{equation}
T(\tau_0, {\bf x_\perp}, \varsigma) = \varepsilon^{-1}\!\left( \varepsilon_0 \frac{\rho(b, {\bf x_\perp}, \varsigma)}{\rho(0, {\bf 0}, 0)} \right) ,
\label{inilambda}
\end{equation}
where $b$ is the impact parameter and the proportionality constant $\varepsilon_0$ is chosen in such a way as to correspond to a given central temperature, $T_0 = \varepsilon^{-1}(\varepsilon_0)$.

The density of sources is constructed using the following mixed model
\begin{equation}
\rho(b, {\bf x_\perp}, \varsigma) \equiv \left[\!\frac{}{} (1 - \kappa) (\rho_{\rm WN}^+(b, {\bf x_\perp}) +\rho_{\rm WN}^-(b, {\bf x_\perp})) + \, 2 \,\kappa\, \rho_{\rm BC}(b, {\bf x_\perp}) \frac{}{}\!\right] \rho_L(\varsigma - \varsigma_S({b, \bf x_\perp})) \, ,
\label{sources}
\end{equation}
where $\rho^{\pm}_{\rm WN}$ is the density of wounded nucleons from the left/right-moving nuclei and $\rho_{\rm BC}$ is the density of binary collisions, both of which are obtained using the optical limit of the Glauber model
\begin{equation}
\rho_{\rm WN}^\pm(b, {\bf x_\perp})  \equiv T\left( {\bf x_\perp}\!\mp\!\frac{{\bf b_\perp}}{2}\right)\!\left[ 1\! -\! e^{- \sigma_{in} T\left( {\bf x_\perp} \pm  \frac{{\bf b_\perp}}{2}\right)} \right] ;
\;\;\;\;\;\;
\rho_{\rm BC}(b, {\bf x_\perp})  \equiv \sigma_{in}T\left( {\bf x_\perp}\!+\! \frac{{\bf b_\perp}}{2}\right) T\left( {\bf x_\perp}\!-\!\frac{{\bf b_\perp}}{2}\right) .
\label{sourcesGlauber}
\end{equation}
The longitudinal profile is taken to be
\begin{equation}
\rho_L(\varsigma) \equiv \exp \left[ - \frac{(\varsigma - \Delta \varsigma)^2}{2 \sigma_\varsigma^2} \Theta (|\varsigma| - \Delta \varsigma) \right] .
\label{longprof}
\end{equation}

For the LHC case studied here, we use $\kappa = 0.145$ for the mixing factor and an inelastic cross-section of $\sigma_{ in} = 62$ mb. The parameters in the longitudinal profile (\ref{longprof}) were fitted to reproduce the pseudorapidity distribution of charged particles, with the results being $\Delta\varsigma = 2.5$ and $\sigma_{\varsigma} = 1.4$ and the shift in rapidity was calculated according to the formula \cite{Bozek:2009ty}
\begin{equation}
\varsigma_S \equiv \frac{1}{2}\ln \frac{\rho_{\rm WN}^++ \rho_{\rm WN}^- + v_P (\rho_{\rm WN}^+- \rho_{\rm WN}^-)}{\rho_{\rm WN}^++ \rho_{\rm WN}^- - v_P (\rho_{\rm WN}^+- \rho_{\rm WN}^-)} \, ,
\label{shift}
\end{equation}
where all functions are understood to be evaluated at a particular value of $b$ and ${\bf x_\perp}$. The participant velocity is defined as $v_P \equiv \sqrt{(\!\sqrt{s}/2)^2-(m_N/2)^2}/(\!\sqrt{s}/2)$, $m_N$ is the nucleon mass, and $\sqrt{s}$ is the center-of-momentum collision energy. In Eq.~(\ref{sourcesGlauber}) we have made use of the thickness function
\begin{equation}
T({\bf x_\perp}) \equiv \int dz \; \rho_{\rm WS}({\bf x_\perp},z) \, ,
\label{thickness}
\end{equation}
where the nuclear density is given by the Woods-Saxon profile
\begin{equation}
\rho_{\rm WS}({\bf x_\perp},z) \equiv \rho_0 \left[ 1 + \exp\left(\frac{\sqrt{{\bf x_\perp}^2 + z^2} - R}{a}\right)\right]^{-1}.
\label{WS}
\end{equation}
For Pb-Pb collisions, we use $\rho_0 = 0.17\, {\rm fm}^{-3}$ for the nuclear saturation density, $R = 6.48$ fm for the nuclear radius, and $a = 0.535$ fm for the surface diffuseness of the nucleus. 

In the calculations presented herein, we assumed that the produced matter has no initial transverse flow, i.e. $u_x (\tau_0, {\bf x_\perp}, \varsigma) = u_x (\tau_0, {\bf x_\perp}, \varsigma) = 0$, while the initial longitudinal flow is of Bjorken form $u_z(\tau_0, {\bf x_\perp}, \varsigma) = z/t$. For simplicity, in this work the initial anisotropy parameters are assumed to be homogeneous and isotropic, $\vec{\alpha}(\tau_0, {\bf x_\perp}, \varsigma) = 1$.

\begin{table}
    \centering
    \begin{minipage}{.5\textwidth}

\centering        
\begin{tabular}{||c c c c c||} 
 \hline
$c_{\rm min}$ & $c_{\rm max}$  & $b_{\rm min}$  & $b_{\rm max}$  & $\langle b \rangle$ \\ [0.5ex] 
 \hline\hline
0. & 0.05 & 0 & 3.473 & 2.315\\ 
 \hline 
0.05 & 0.1 & 3.473 & 4.912 & 4.234\\ 
 \hline 
0.1 & 0.2 & 4.912 & 6.946 & 5.987\\ 
 \hline 
0.2 & 0.3 & 6.946 & 8.507 & 7.753\\ 
 \hline 
0.3 & 0.4 & 8.507 & 9.823 & 9.181\\ 
 \hline 
0.4 & 0.5 & 9.823 & 10.983 & 10.414\\ 
 \hline 
\end{tabular}

    \end{minipage}%
    \begin{minipage}{0.5\textwidth}

\vspace{-4.4mm}
 \centering
 \begin{tabular}{||c c c c c||} 
 \hline
$c_{\rm min}$ & $c_{\rm max}$  & $b_{\rm min}$  & $b_{\rm max}$  & $\langle b \rangle$ \\ [0.5ex] 
 \hline\hline
0.5 & 0.6 & 10.983 & 12.031 & 11.515\\ 
 \hline 
0.6 & 0.7 & 12.031 & 12.995 & 12.519\\ 
 \hline 
0.7 & 0.8 & 12.995 & 13.893 & 13.449\\ 
 \hline 
0.8 & 0.9 & 13.893 & 14.795 & 14.334\\ 
 \hline 
0.9 & 1. & 14.795 & 20 & 15.608\\ 
 \hline 
\end{tabular}

    \end{minipage}
\caption{The optical Glauber relationship used herein between the centrality class ($c_{\rm min}$, $c_{\rm max}$), the minimum and maximum impact parameter in that class ($b_{\rm min}$, $b_{\rm max}$), and the average impact parameter $\langle b \rangle$.}
\label{glaubertable}
\end{table}

\subsection{Numerical methods}
\label{num}

We solve the 3+1d aHydro equations on a lattice with spacing $a_\perp = 0.2$ fm and $a_\zeta = 0.2$ with $N_x = N_y = N_\varsigma = 128$.  For temporal updates we used 4th-order Runge-Kutta with a temporal step size of \mbox{$\epsilon=0.01$ fm/c}.  A weighted LAX scheme with a rather small weighting factor of $\lambda = 0.01$ was used to regulate possible numerical instabilities associated with shock-wave formation~\cite{Martinez:2012tu}.  We varied the centrality of the collision by setting the impact parameter to the average value expected in each centrality class according to the optical Glauber model as indicated in Table \ref{glaubertable}.  Each of the configurations was evolved until the maximum effective temperature in the entire volume was below 120 MeV.  From each of these results, a freeze-out hypersurface corresponding to a fixed effective temperature was extracted.  The microscopic parameters $\vec{\alpha}$, $\vec{u}$, and $\lambda$ on the hypersurface were then fed to a version of THERMINATOR 2 which had been modified to account for an ellipsoidal distribution function \cite{Chojnacki:2011hb}.  THERMINATOR 2 produces sampled event-by-event hadronic production from the exported freezeout hypersurface and also performs hadronic feed down (resonance decays) for each sampled event.

\section{Results}
\label{results}

\begin{figure}[t]
\centerline{
\includegraphics[width=0.5\linewidth]{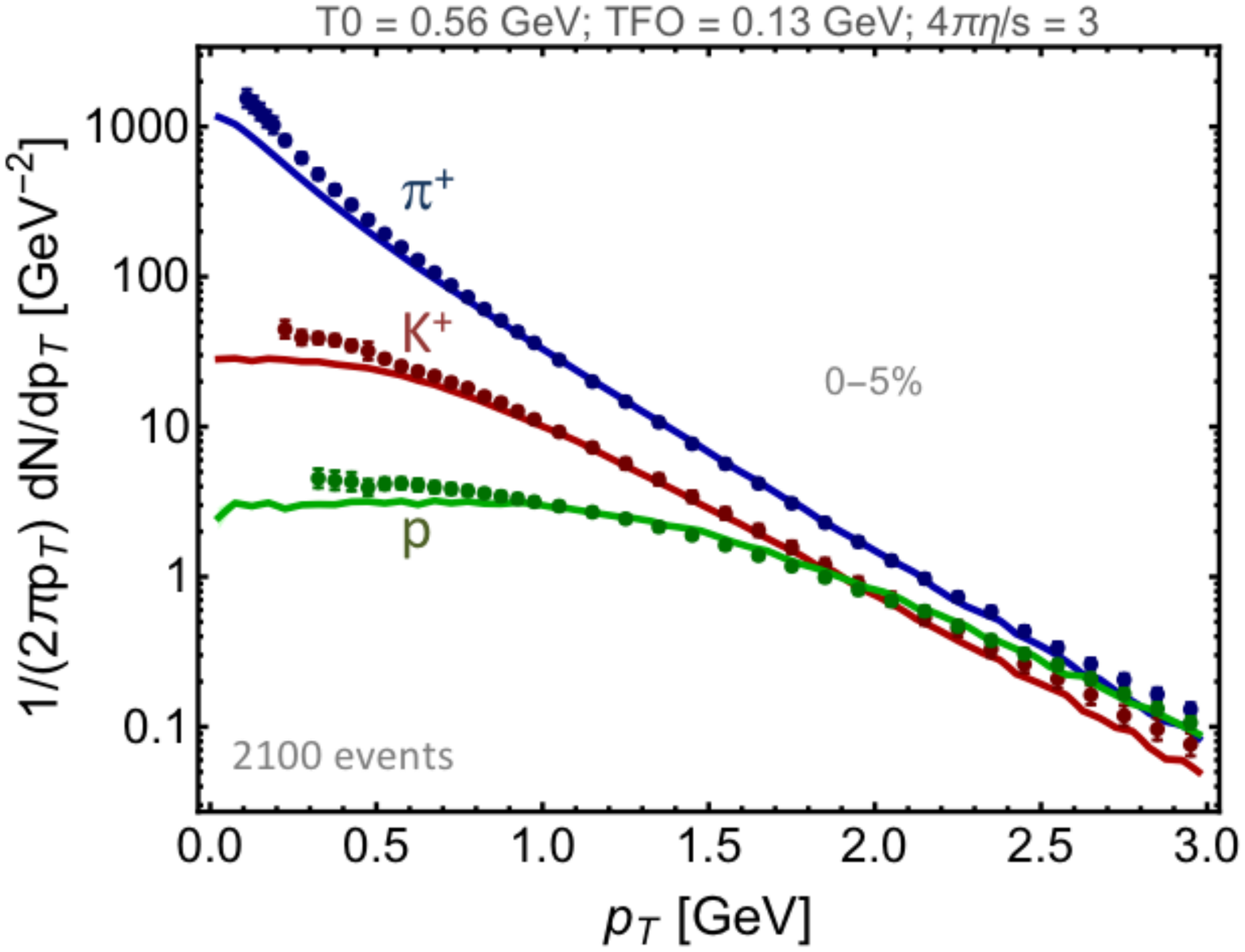}
\includegraphics[width=0.48\linewidth]{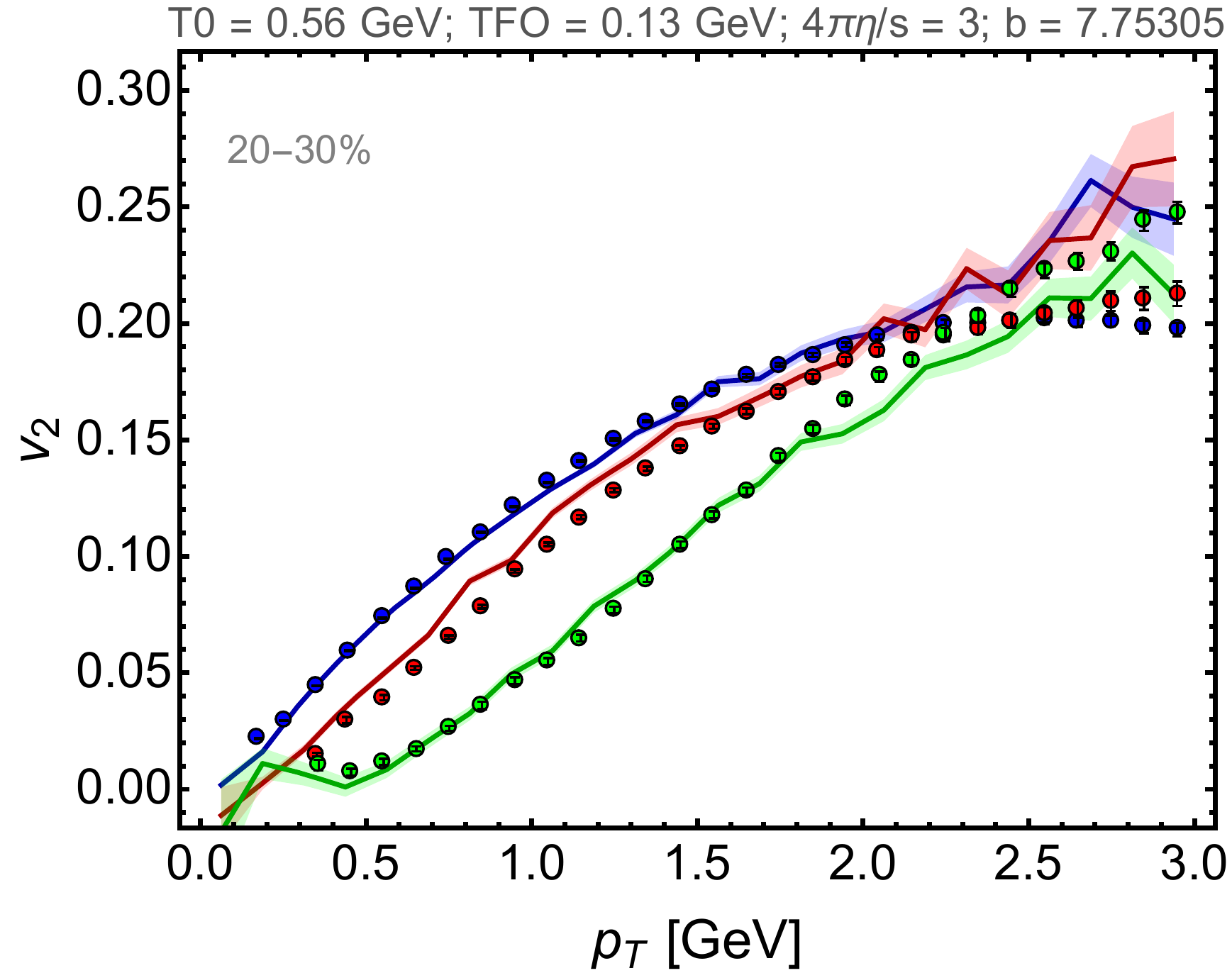}
}
\caption{Comparison of aHydro model predictions with experimental data for $\pi^+$ (blue), $K^+$ (red), and $p$ production (green). The left panel shows the particle spectra in the 0-5\% centrality class and the right panel shows $v_2$ in the 20-30\% centrality class, both as a function of transverse momentum $p_T$. The best fit corresponded to $T_0=$ 0.56 GeV, $T_{\rm FO}=$ 0.13 GeV, and $4\pi\eta/s=3$. Data shown are from the ALICE collaboration~\cite{Abelev:2013vea,Abelev:2014pua}. Experimental error bars shown are statistical only.}
\label{fig1}
\end{figure}

Herein we consider LHC 2.76 TeV Pb-Pb collisions and compare to experimental data available from the ALICE collaboration~\cite{Abelev:2013vea,Abelev:2014pua}.
Using the initial conditions specified previously, we varied \mbox{$4\pi\eta/s \in \{1,2,3\}$} and, in each case, varied the initial central temperature $T_0$ and the freeze-out temperature $T_{\rm FO}$ and compared with ALICE data.  From our preliminary analysis, the values of \mbox{$T_0 \simeq 560$ MeV} and $T_{\rm FO} \simeq 130$ MeV gave the best fit to the spectrum and $v_2$.  

In Fig.~\ref{fig1} we show the resulting $\pi^+$, $K^+$, and $p$  transverse momentum spectrum in the 0-5\% centrality class (left) and $v_2$ in the 20-30\% centrality class (right).  As can be seen from this figure, the model does a very good job reproducing the identified particle $v_2$ in the 20-30\% centrality class; however, we see significant deviations in the low-$p_T$  $\pi^+$, $K^+$, and $p$ spectra.  This discrepancy is most likely a result of the ``standard method'' for implementing a realistic equation of state.  As was shown in Ref.~\cite{Alqahtani:2016rth}, one finds that the standard method significantly underestimates the number of low $p_T$ hadrons, whereas the quasiparticle method for implementing the equation of state is in better agreement with standard second-order viscous hydrodynamics at low $p_T$.

\begin{figure}[t]
\centerline{
\includegraphics[width=0.48\linewidth]{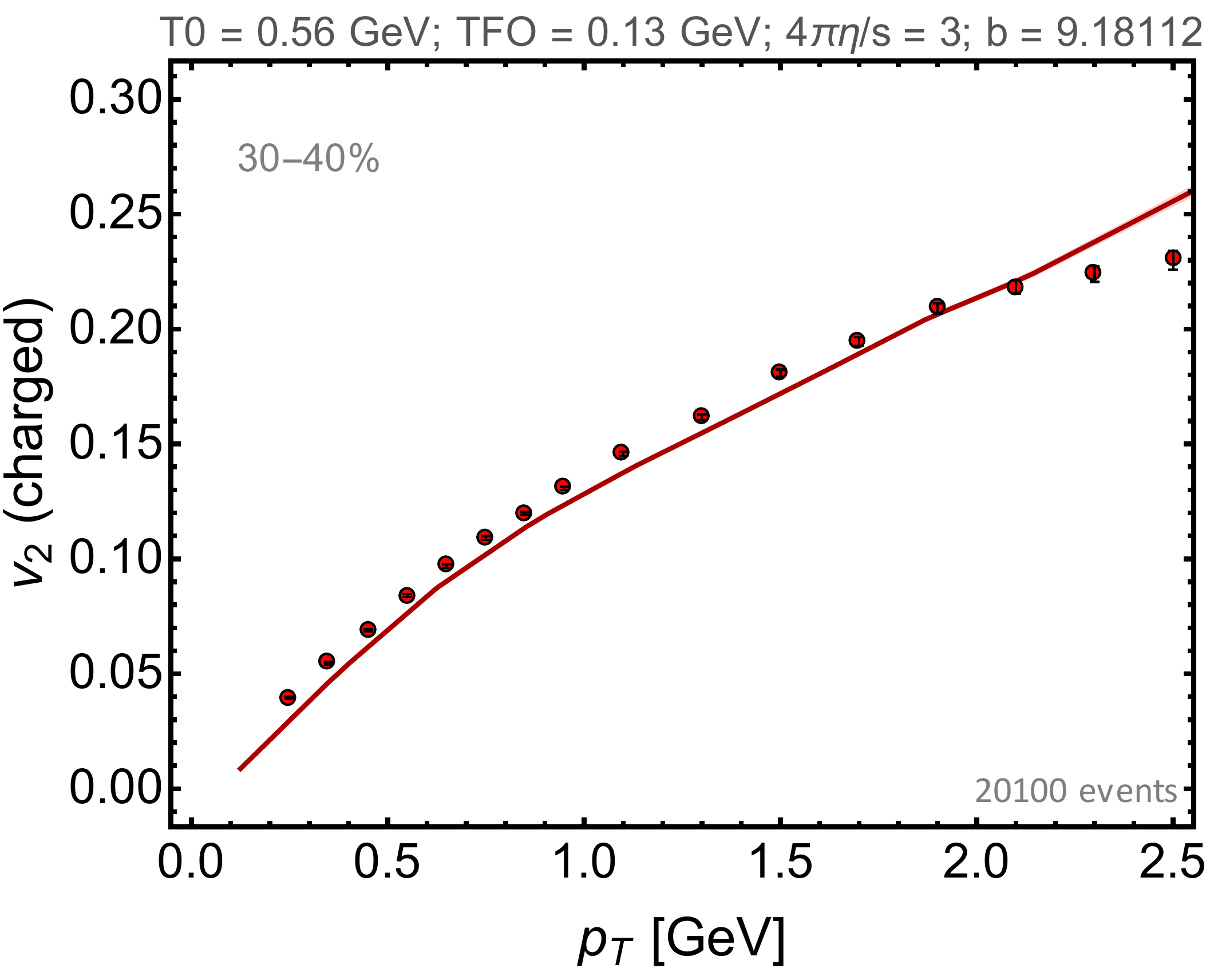}
\includegraphics[width=0.5\linewidth]{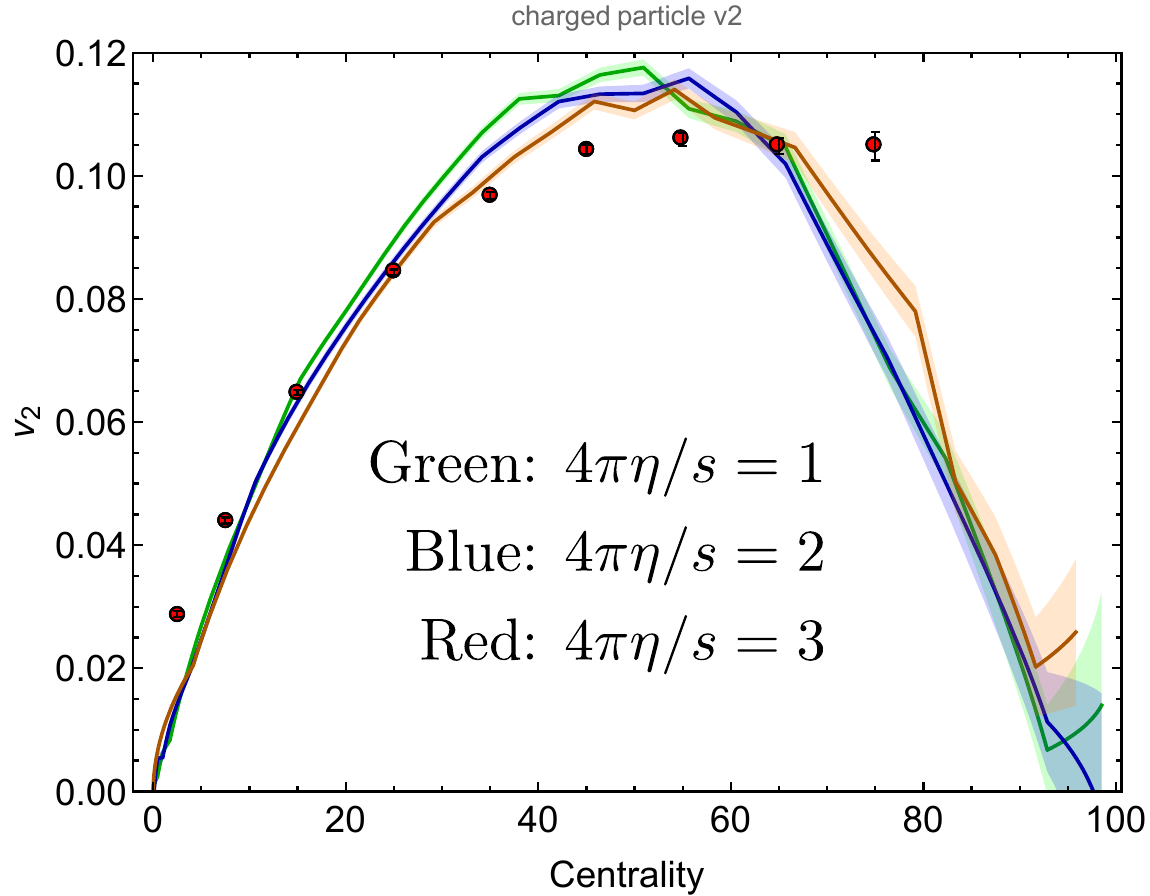}
}
\caption{Comparison of aHydro model predictions with experimentally measured charged particle $v_2$ as a function of $p_T$ in the 30-40\% centrality class (left) and the $p_T$-integrated charged particle $v_2$ as a function of centrality (right). Data shown are from the ALICE collaboration~\cite{Abelev:2013vea,Abelev:2014pua}. Experimental error bars shown are statistical only.}
\label{fig2}
\end{figure}

In Fig.~\ref{fig2} we show two results.  In the left panel, we show the charged particle $v_2$ as a function of $p_T$ in the 30-40\% centrality class and on the right we show the $p_T$-integrated charged particle $v_2$ as a function of centrality.  As can be seen from the left panel, there is reasonable agreement between the aHydro model employed herein and the charged particle $v_2$.  In the right panel, we show three results corresponding to \mbox{$4\pi\eta/s \in \{1,2,3\}$} which are indicated in red, blue, and green, respectively, and are ordered from top to bottom at 35\% centrality, respectively.  As we can see from the right panel there is disagreement between the model and the data for very central events.  This is to be expected since we used smooth optical Glauber initial conditions.  All $v_2$ generated in central collisions is due to fluctuations and, in the first two centrality classes, there are significant contributions to $v_2$ from initial state fluctuations.  In the peripheral centrality classes we also note that the model prediction falls more quickly to zero than the data seem to indicate; however, we note there is a large systematic error associated with the reported $v_2$ in the most peripheral centrality class.  Finally, we note that the maximum model sensitivity to the assumed values of $\eta/s$ occurs at around 40\% centrality.

\section{Conclusions}
\label{conclusions}

In this proceedings, I have reviewed recent progress in anisotropic hydrodynamics.  The anisotropic hydrodynamics framework has been checked against exact solution of the Boltzmann equation in some non-trivial but simple cases in which exact solution is possible and one finds that, in all cases considered, the anisotropic hydrodynamics approach most accurately reproduces the evolution of the system.  Building on this success, the current focus of the aHydro program is to turn the formalism into a practical phenomenological tool that can be used to model the non-equilibrium dynamics of the QGP created in AB nuclear collisions including pA and, down the road, pp collisions.  As demonstrated herein, significant progress in this direction has been made.  We now have functioning 3+1d versions of leading-order aHydro which include multiple anisotropy parameters that are associated with the shear and bulk corrections to the distribution function.  

The code now includes ``anisotropic freeze-out'' which is tightly integrated with a customized version of THERMINATOR 2, which takes care of the final hadronic production and resonance decays.  Comparisons with ALICE data indicate that, even given the fact that we are using smooth Glauber initial conditions, the model is able to reproduce the qualitative features of the data overall, but in the case of the $p_T$-differential $v_2$ we find very promising agreement between the 3+1d aHydro results and experimental data.  The key standout phenomenologically lies in our inability to properly fit the low $p_T$ part of the hadronic spectra.  We have suggested that this is due to an inconsistency inherent in the standard method for implementing the equation of state in aHydro.  Looking to the future, we are currently working on optimizing the quasiparticle aHydro approach in order to make it feasible to use in 3+1d simulations.  Initial testing shows that this will be possible and offers hope that one can have a self-consistent way to implement conformal symmetry breaking while at the same time having an efficient code.  This will become critical as we begin explorations of fluctuating initial conditions.

\vspace{1.5cm}

\section*{Acknowledgements}
I thank my collaborators M. Alqahtani, W. Florkowski, M. Nopoush, and R. Ryblewski.
This work was supported by the U.S. Department of Energy, Office of Science, Office of Nuclear Physics under Award No. DE-SC0013470.

\bibliography{STRICKLAND_Michael_CONF12}

\end{document}